\documentclass[12pt]{iopart}

\usepackage{bm}
\usepackage{mathrsfs}
\usepackage{verbatim}
\usepackage[all,cmtip]{xy}
\usepackage{amssymb}
\usepackage{graphics}
\usepackage{amsthm}
\usepackage{color}
\usepackage{amscd}
\usepackage{cases}
\usepackage{amssymb}
\usepackage{epsfig}
\usepackage{color}
\usepackage{dsfont}
\definecolor{darkred}  {rgb}{0.5,0,0}
\definecolor{darkblue} {rgb}{0,0,0.5}
\definecolor{darkgreen}{rgb}{0,0.5,0}
\usepackage{hyperref}
\hypersetup{
	pdftitle = {QRT Proposal},
	pdfauthor = {},
	colorlinks = true,
	urlcolor  = blue,         % color of external links
	linkcolor = red,     % color of internal links
	citecolor = blue,    % color of links to bibliography
	filecolor = darkred       % color of file links
}

\def\ra{\rangle}
\def\la{\langle}
\def\bb{\mathbb}

\def\ra{\rangle}
\def\la{\langle}
\def\bb{\mathbb}

\newtheorem{theorem}{Theorem}

\newtheorem{pro}[theorem]{Proposition}

\theoremstyle{remark}
\newtheorem{remark}{Remark}

\newcommand{\beax}{\begin{eqnarray*}}
\newcommand{\eeax}{\end{eqnarray*}}

\def\be{\begin{eqnarray}}
\def\ee{\end{eqnarray}}
\newcommand{\bea}{\begin{eqnarray}}
\newcommand{\eea}{\end{eqnarray}}

\newcommand{\mF}{\mathcal{F}}
\newcommand{\mE}{\mathcal{E}}

\newcommand{\mH}{\mathcal{H}}

\newcommand{\mS}{\mathcal{S}}

\newcommand{\lr}{\rangle\langle}

\def\>{\rangle}
\def\<{\langle}
\def\tr{ \mathrm{Tr}}

\begin{document}

\title[Genuine multipartite entanglement measure]{Genuine multipartite entanglement measure}

\author{Yu Guo}

\address{Institute of Quantum Information Science, School of Mathematics and Statistics, Shanxi Datong University, Datong, Shanxi 037009, China}
\ead{guoyu3@aliyun.com}

\author{Yanping Jia}
%\email{jyp1982328@126.com}
\address{Institute of Quantum Information Science, School of Mathematics and Statistics, Shanxi Datong University, Datong, Shanxi 037009, China}

\author{Xinping Li}
%\email{lixp0802@163.com}
\address{Institute of Theoretical Physics, Shanxi Datong University, Datong, Shanxi 037009, China}

\author{Lizhong Huang}

\address{Institute of Quantum Information Science, School of Mathematics and Statistics, Shanxi Datong University, Datong, Shanxi 037009, China}
%\ead{guoyu3@aliyun.com}

\begin{abstract}
Quantifying genuine entanglement is a crucial task in quantum information theory.
In this work, we give an approach of 
constituting genuine $m$-partite entanglement measures from any bipartite entanglement
and any $k$-partite entanglement measure, $3\leq k<m$. 
In addition, as a complement to the three-qubit concurrence triangle proposed in [Phys. Rev. Lett., 127, 040403], we show that the triangle relation is also valid for any continuous entanglement measure and system with any dimension. 
We also discuss the tetrahedron structure for the four-partite system via the triangle relation associated with tripartite and bipartite entanglement respectively. For multipartite system that contains more than four parties, there is no symmetric geometric structure as that of tri- and four-partite cases.

\vspace{2pc}
\end{abstract}

%\begin{indented}
%	\item[]January 2018
%\end{indented}
%\pacs{03.65.Ud, 03.65.Db, 03.67.Mn.}

\maketitle

\section{Introduction}

Entanglement as one of the most puzzling features
of quantum mechanics has been recognized as a crucial resource in
many quantum information processing protocols in the last three decades~\cite{Nielsen,Bennett1996prl,Horodecki2009,Guhne2009,Burkhart2021prx,Yuxiaodong2021pral}.
Numerous entanglement measures, especially for the bipartite entanglement, have been proposed
so far~\cite{Horodecki2009,Guhne2009,Plenio2007,Ma2011,Hong2012pra,Gao2014prl,Hong2012pra,Gao2010pra,Yang2008prl,Szalay,Liming2017pra,G2020}.
Yet, the situation
is much more complicated for multipartite systems.
How to quantify the genuine entanglement contained in the multipartite state is still
lacking in research.

The first effort in this connection is the ``residual tangle'' which reports the genuine three-qubit
entanglement by Coffman \textit{et al}. in 2000~\cite{Coffman}.
In 2011, Ma \textit{et al}.~\cite{Ma2011} established some conditions for a quantity to be
a genuine multipartite entanglement measure (GMEM) and proposed a GMEM, called genuinely multipartite concurrence (GMC), according to the origin bipartite concurrence.
In such a sense, the ``residual tangle'', the generalizations of concurrence~\cite{Hong2012pra,Hiesmayr2008pra,Carvalho2004prl}, 
the generalization of negativity~\cite{Jungnitsch2011prl}
the SL-invariant 
multipartite measure of entanglement~\cite{Verstraete2003pra,Luque2003pra,Osterloh2005pra,Osterloh2009jmp,Gour2010prl,Viehmann2011pra}, 
and the $\alpha$-entanglement entropy~\cite{Szalay} are not GMEMs since all these quantities
do not obey all the genuine entanglement conditions in Ref.~\cite{Ma2011}.
Very recently, a new genuine three-qubit entanglement measure, called \textit{concurrence triangle}, was put forward in Ref.~\cite{Xie2021prl}. It is quantified as the square root of the area of concurrence triangle.
Under this scenario, the GHZ state is more entangled than the W state.

Apart from GMC and the concurrence triangle, there are very few GMEM by now. We list here for reader's convenience.
The GMC is further explored in Ref.~\cite{Rafsanjani},
the generalized geometric measure is introduced in Refs.~\cite{Sen2010pra,Sadhukhan2017pra},
the average of ``residual tangle'' and GMC~\cite{Emary2004pra} are shown to be GMEMs.
Another one is the divergence-based GMEM presented in~\cite{Contreras-Tejada,Das}.
The aim of this work is to construct (genuine) $m$-partite entanglement measures
from bipartite entanglement measures and $k$-partite entanglement measures, $3\leq k<m$,
and, in addition,
generalize the concurrence triangle in Ref.~\cite{Xie2021prl}
to any other continuous entanglement measure and system with any dimension.

The rest of this article is organized as follows.
In Section II we review some necessary concepts.
Section III discusses the triangle relation and present genuine tripartite entanglement measure via bipartite entanglement measure.
In Section IV we establish genuine four-partite entanglement measures via the tripartite entanglement measure,
bipartite entanglement measure and the genuine tripartite entanglement measure defined in Section III, respectively. 
The approach for extending to general multipartite case is also discussed.
In Section V, we investigate the tetrahedron structure for any four-partite state by tripartite and bipartite entanglement measures, respectively.
A summary is conclude in the last section. 
For simplicity, we fix some notations. Throughout this paper, we denote by $\mH^{A_1A_2\cdots A_m}:=\mathcal{H}^{A_1}\otimes
\mathcal{H}^{A_2}\otimes\cdots\otimes\mathcal{H}^{A_m}$
an $m$-partite Hilbert space with finite dimension and by $\mS^{X}$ we denote the set of density
operators acting on $\mH^{X}$.

\section{Preliminary}

For convenience, in this section, we recall the concepts of genuine entanglement, the bipartite entanglement measure,
the complete multipartite entanglement measure and the GMEM in detail.

\subsection{Multipartite entanglement}

An $m$-partite pure state $|\psi\rangle\in \mathcal{H}^{A_1A_2\cdots A_m}$ is called biseparable if it can be written as
%\begin{equation}\label{}
$|\psi\rangle=|\psi\rangle^X \otimes |\psi\rangle^Y$
for some bipartition of $A_1A_2\cdots A_m$ (for example, for a 4-partite state, $A_1A_3|A_2A_4$ is a bipartition of $A_1A_2A_3A_4$). 
$|\psi\ra$ is said to be $k$-separable if
$|\psi\ra=|\psi\ra^{X_1}|\psi\ra^{X_2}\cdots|\psi\ra^{X_k}$
for some $k$-partition of $A_1A_2\cdots A_m$ (for example, for a 5-partite state, $A_1A_3|A_2|A_4A_5$ is a 3-partition of $A_1A_2A_3A_4A_5$ with $X_1=A_1A_3$, $X_2=A_2$ and $X_3=A_4A_5$).
$|\psi\ra$ is called fully separable if it is $m$-separable.
An $m$-partite mixed state $\rho$ is
biseparable if it can be written as a convex combination of
biseparable pure states
%\begin{equation}\label{}
$\rho=\sum_{i}p_i|\psi_i\rangle \langle\psi_i|$, 
%\end{equation}
wherein the contained $\{|\psi_i\rangle\}$ can be biseparable with respect to different
bipartitions (i.e., a mixed biseparable state does not need to be separable with respect to any particular bipartition). If $\rho$ is not biseparable, then it is
called genuinely entangled.
$\rho$ is said to be $k$-separable with respect to partition $X_1|X_2|\cdots|X_k$
if $\rho=\sum_{i}p_i|\psi_i\rangle \langle\psi_i|$
with $|\psi_i\ra$ is $k$-separable with respect to partition $X_1|X_2|\cdots|X_k$ for
any $i$.
Throughout this paper, for any $\rho\in\mS^{A_1A_2\cdots A_m}$ and any given $k$-partition $X_1|X_2|\cdots|X_k$
of $A_1A_2\cdots A_m$,
we denote by $\rho^{X_1|X_2|\cdots|X_k}$ the state for which we consider it as a $k$-partite state
with respect to the partition $X_1|X_2|\cdots|X_k$.

It is clear that whenever a
state is $k$-separable, it is automatically also $l$-separable for all $1<l<k$. We denote the set of all $k$-separable states by $\mS_k$ ($k=2,3,\cdots,m$), then $\mS_m\subsetneq \mS_{m-1}\subsetneq\cdots \subsetneq \mS_2\subsetneq \mS$. $\mS\backslash\mS_k$ is the set of all $k$-entangled states. $\mS\backslash\mS_2$ is just the set of all genuinely entangled (or equivalently, 2-entangled) states (here we denote $\mS^{A_1A_2\cdots A_m}$ by $\mS$ for simplicity).
If a
state is $k$-entangled, it is automatically also $j$-entangled for all $k<j\leqslant m$ but not vice versa.

\subsection{Bipartite entanglement measure}

A function $E:
\mS^{AB}\to\bb{R}_{+}$ is called an \emph{entanglement measure} if
it satisfies~\cite{Vedral97}: 
\begin{itemize}
	\item (E-1) $E(\rho)=0$ if $\rho$ is
	separable; 
	\item (E-2) $E$ cannot increase under local operation and classical communications (LOCC), i.e.,
	$E(\varepsilon(\rho))\leqslant E(\rho)$ for any LOCC $\varepsilon$ [(E-2) 
	implies that $E$ is invariant under local unitary
	operations, i.e., $E(\rho)=E(U^A\otimes U^B\rho U^{A,\dag}\otimes
	U^{B,\dag})$ for any local unitaries $U^A$ and $U^B$].
	The map $\varepsilon$ is completely positive and trace preserving 
	(CPTP). 
\end{itemize}
In general, LOCC can be stochastic, 
in the sense that $\rho$ can be converted to $\sigma_{j}$ with some probability $p_j$. 
In this case, the map from $\rho$ to $\sigma_{j}$ can not be described in general by a CPTP map. 
However, by introducing a ``flag'' system $A'$, we can view the ensemble $\{\sigma_{j},p_j\}$ 
as a classical quantum state $\sigma':=\sum_{j}p_j|j\lr j|^{A'}\otimes\sigma_{j}$. 
Hence, if $\rho$ can be converted by LOCC to $\sigma_{j}$ with probability $p_j$, 
then there exists a CPTP LOCC map $\Phi$ such that $\Phi(\rho)=\sigma'$.
Therefore, the definition above of a measure of entanglement captures also probabilistic transformations. 
In particular, $E$ must satisfy $E\left(\sigma'\right)\leqslant E\left(\rho\right)$.

Any measure of entanglement studied in literature satisfy
\beax
E\left(\sigma'\right)=\sum_{j}p_jE(\sigma_{j})\;,
\eeax
which is very intuitive since $A'$ is just a classical system encoding the value of $j$. 
In this case the condition $E\left(\sigma'\right)\leqslant E\left(\rho\right)$ 
becomes 
\bea\label{average}
\sum_{j}p_jE(\sigma_{j})\leqslant E\left(\rho\right).
\eea That is, LOCC can 
not increase entanglement on average (it is possible that $E(\sigma_{j_0})>E(\rho)$ for some $j_0$). 
An entanglement measure $E$ is said to be an entanglement
monotone~\cite{Vidal2000} if it satisfies Eq.~(\ref{average}) and is convex additionally (almost all entanglement measures are entanglement monotones although not all, see in Ref.~\cite{Plenio2005} for an example, the logarithmic negativity is not convex).

\subsection{Complete multipartite entanglement measure}

A function 
$E^{(m)}: \mS^{A_1A_2\cdots A_m}\to\bb{R}_{+}$ 
is called an $m$-partite entanglement measure~\cite{Horodecki2009,Hong2012pra,Hiesmayr2008pra} if
it satisfies:
\begin{itemize}
	\item {\bf(E1)} $E^{(m)}(\rho)=0$ if $\rho$ is fully separable;
	\item {\bf(E2)}
	$E^{(m)}$ cannot increase
	under $m$-partite LOCC.
\end{itemize}
An $m$-partite entanglement measure $E^{(m)}$ is said to be 
an $m$-partite entanglement monotone if it is convex and 
does not increase on average under $m$-partite stochastic LOCC.
$E^{(m)}$ is called a {unified}
multipartite entanglement measure if it also satisfies the following
condition~\cite{G2020}: 
\begin{itemize}
	\item {\bf(E3)}~\textit{the unification condition}, i.e., 
	$E^{(m)}$ is consistent with $E^{(k)}$ for any $2\leqslant k<m$.
\end{itemize}
The unification condition should be comprehended in the following sense~\cite{G2020} (We take $m=3$ for example).
Let $|\psi\ra^{ABC}$ be a bi-separable pure state in $\mH^{ABC}$, e.g., 
$|\psi\ra^{ABC}=|\psi\ra^{AB}|\psi\ra^{C}$ then
\beax
E^{(3)}(|\psi\ra^{AB}|\psi\ra^{C})=E^{(2)}(|\psi\ra^{AB}).
\eeax
In this way, the link between
$E^{(2)}$ and $E^{(3)}$ can be established. 
The unification condition also requires the
measure must be 
\textit{invariant under the permutations of the subsystems}, i.e.,  
\beax
E^{(3)}(\rho^{ABC})=E^{(3)}(\rho^{\pi(ABC)}),\quad \textrm{for any}~\rho^{ABC}\in\mS^{ABC},
\eeax 
where $\pi$ is a permutation of the subsystems. 
In addition, 
\beax
E^{(3)}(\rho^{ABC})\geqslant E^{(2)}(\rho^{XY}),\quad \rho^{XY}=\tr_{Z}(\rho^{ABC}), ~X,Y,\in\{A,B,C\}
\eeax
for any $\rho^{ABC}\in\mS^{ABC}$. 
That is, a unified multipartite entanglement measure $E^{(m)}$ is indeed a series of
measures $\{E^{(m)}, E^{(m-1)},\dots, E^{(2)}\}$.

$E^{(m)}$ is called a {complete}
multipartite entanglement measure if it satisfies both {\bf(E3)} above and the following~\cite{G2020}: 
\begin{itemize}
	\item {\bf(E4)}~$E^{(m)}(\rho^{A_1A_2\cdots A_m})\geqslant
	E^{(k)}(\rho^{A_1'A_2'\cdots A_k'})$
	holds for all $\rho^{A_1A_2\cdots A_m}\in\mS^{A_1A_2\cdots A_m}$, $A_1'A_2'\cdots A_k'$ 
	is any $k$-partition of $A_1A_2\cdots A_m$.
\end{itemize}

For instance, for any pure state $|\psi\ra\in\mH^{AB}$, the entanglement of formation 
(EoF)~\cite{Bennett1996pra,Horodecki01}, $E_f$, is defined as
\beax
E_f^{(2)}(|\psi\ra)=E_f(|\psi\ra)=S(\rho^A)=S(\rho^B)
=\frac12\left[S(\rho^A) +S(\rho^B)\right]
\eeax
and the tangle is defined by~\cite{Rungta2003pra}
\beax
\tau(|\psi\ra)&=& 2\left(1- {\rm Tr}\left( \rho^A\right)^2\right) =
2-{\rm Tr}\left(  \rho^A\right)^2-{\rm Tr}\left(  \rho^B\right)^2.
\eeax
These measures are extended into three-partite case by~\cite{G2020}
\beax
E_f^{(3)}\left( |\psi\rangle\right):=\frac12\left[S(\rho^A)+S(\rho^B)+S(\rho^C) \right],\\
\tau^{(3)}(|\psi\ra):=3- {\rm Tr}\left( \rho^A\right) ^2-{\rm Tr}\left( \rho^B\right) ^2-\tr\left( \rho^C\right) ^2,
\eeax
for pure state $|\psi\ra\in\mH^{ABC}$.
By the convex-roof extension, the measures can be defined for mixes states.
In Ref.~\cite{G2020}, we show that 
$\{E_f^{(3)}, E_f^{(2)},\}$ and $\{\tau^{(3)}, \tau^{(2)}\}$
are complete tripartite entanglement measures (one can see other complete tripartite entanglement measures in Ref.~\cite{G2020} for detail).

\subsection{Genuine entanglement measure}

A function $E_{g}: \mS^{A_1A_2\cdots A_m}\to\bb{R}_{+}$ is defined to be a GMEM if it admits the following conditions~\cite{Ma2011}:
\begin{itemize}
	\item {\textbf{(GE1)}}~$E_{g}(\rho)=0$ for any biseparable $\rho\in\mS^{A_1A_2\cdots A_m}$.
	\item {\textbf{(GE2)}}~ $E_{g}(\rho)>0$ for any genuinely entangled state $\rho\in \mathcal{S}^{A_1A_2\cdots A_m}$.(This item can be weakened as: $E_{g}(\rho)\geqslant 0$ for any genuinely entangled state $\rho\in \mathcal{S}^{A_1A_2\cdots A_m}$. That is, maybe there exists some state which is genuinely entangled such that $E_{g}(\rho)= 0$. In such a case, the measure is called not faithful. Otherwise, it is called faithful. For example, the ``residual tangle'' is not faithful since it is vanished for the $W$ state.)
	\item {\textbf{(GE3)}}~ $E_{g}(\sum_ip_i\rho_i)\leqslant \sum_ip_iE_g(\rho_i)$ for any $\{p_i,\rho_i\}$, $\rho_i\in\mathcal{S}^{A_1A_2\cdots A_m}$, $p_i>0$, $\sum_ip_i=1$.
	\item {\textbf{(GE4)}}~ $E_{g}(\rho)\geqslant E_g(\rho')$ for any $m$-partite LOCC $\varepsilon$, $\varepsilon(\rho)=\rho'$.	
\end{itemize}
Note that \textbf{(GE4)} implies $E_{g}$ is invariant under local unitary transformations.
$E_{g}$ is said to be 
a genuine multipartite entanglement monotone if it  
does not increase on average under $m$-partite stochastic LOCC.

We recall a genuine measure, concurrence fill, defined in Ref.~\cite{Xie2021prl}.
The concurrence triangle for three-qubit state is constructed from the following
triangle relation~\cite{Qian2018pra,Zhu2015pra,Xie2021prl}:
\bea
C^2_{A|BC}\leqslant C^2_{AB|C}+C^2_{B|AC}.
\eea
That is, for any three-qubit pure state $|\psi\ra^{ABC}$,
edges with $C^2_{A|BC}$, $C^2_{AB|C}$ and $C^2_{B|AC}$
make up a triangle. Then the square root of the area of this triangle, i.e., the so-called concurrence fill, quantifies the amount of genuine entanglement 
contained in the state.
It can then be defined for mixed state via the convex-roof extension.
(We note here that, whether the concurrence fill is monotonic under LOCC is not proved in Ref.~\cite{Xie2021prl}.
We call it genuine entanglement measure throughout this paper.)

Similarly, we call a function $E_{m(k)}: \mS^{A_1A_2\cdots A_m}\to\bb{R}_{+}$ a $k$-entanglement measure if it admits the following conditions:
\begin{itemize}
	\item {\textbf{($k$-E1)}}~$E_{m(k)}(\rho)=0$ for any $k$-separable $\rho\in\mS^{A_1A_2\cdots A_m}$.
	\item {\textbf{($k$-E2)}}~ $E_{m(k)}(\rho)>0$ for any $k$-entangled state $\rho\in \mathcal{S}^{A_1A_2\cdots A_m}$. (This item can be weakened as: $E_{m(k)}(\rho)\geqslant 0$ for any $k$-entangled state $\rho\in \mathcal{S}^{A_1A_2\cdots A_m}$. That is, maybe there exists some $k$-entangled state such that $E_{m(k)}(\rho)= 0$. In such a case, the measure is called not faithful. Otherwise, it is called faithful.)
	\item {\textbf{($k$-E3)}}~ $E_{m(k)}(\sum_ip_i\rho_i)\leqslant \sum_ip_iE(\rho_i)$ for any $\{p_i,\rho_i\}$, $\rho_i\in\mathcal{S}^{A_1A_2\cdots A_m}$, $p_i>0$, $\sum_ip_i=1$.
	\item {\textbf{($k$-E4)}}~ $E_{m(k)}(\rho)\geqslant E_{m(k)}(\rho')$ for any $m$-partite LOCC $\varepsilon$, $\varepsilon(\rho)=\rho'$.	
\end{itemize}

For simplicity,
throughout this paper, if $E$ is an entanglement measure (bipartite, or multipartite, or $k$-entanglement) for pure states,
we define
\bea\label{eofmin}
E_F(\rho):=\min\sum_ip_iE^{(m)}(|\psi_i\ra)
\eea
and call it the entanglement of formation associated with
$E$, where the minimum is taken over all pure-state
decomposition $\{p_i,|\psi_i\ra\}$ of $\rho$ (namely, the convex-roof extension of $E$. Sometimes, we use $E^F$ to denote $E_F$ hereafter).

\section{Triangle measure}

Motivated by the concurrence triangle,
we firstly consider whether there exists a triangle for any tripartite pure state $|\psi\ra^{ABC}\in\mH^{ABC}$
with arbitrarily given bipartite entanglement measure $E$.
That is, for any given bipartite entanglement measure $E$, whether $E(|\psi\ra^{A|BC})$, $E(|\psi\ra^{B|AC})$ and $E(|\psi\ra^{AB|C})$ can represent the lengths of the 
three edges of a triangle.

\begin{theorem}\label{plogamy-3}
	Let $E$ be a continuous bipartite entanglement measure.
	Then there exists
	$0<\alpha<\infty$ such that
	\bea\label{power1}
	E^{\alpha}(|\psi\ra^{A|BC})\leqslant  E^{\alpha}(|\psi\ra^{B|AC})
	+ E^{\alpha}(|\psi\ra^{AB|C})
	\eea
	for all pure states $|\psi\ra^{ABC}\in\mathcal{H}^{ABC}$ with fixed $\dim\mH^{ABC}=d<\infty$.
\end{theorem}

\begin{proof}
	For any given $|\psi\ra^{ABC}$, we assume that $E(|\psi\ra^{A|BC})=x$, $E(|\psi\ra^{B|AC})=y$
	and $E(|\psi\ra^{AB|C})=z$. 
	If $x\leqslant y$ or $x\leqslant z$, Eq.~(\ref{power1}) is obvious.
	If $x>\max\{y,z\}$, we claim that $y>0$ and $z>0$ (note that, if $x>\max\{y,z\}$, then $\max\{y,z\}>0$).
	In order to see this, we assume with no loss of generality that
	$x>y\geqslant z\geqslant 0$. We now suppose $z=0$, then we
	let 
	\beax
	|\psi\ra^{ABC}=|\psi\ra^{AB}|\psi\ra^C
	\eeax
	for some $|\psi\ra^{AB}\in\mH^{AB}$ and $|\psi\ra^C\in\mH^C$.
	It follows that
	\beax
	E^{A|BC}(|\psi\ra^{ABC})
	=E^{B|AC}(|\psi\ra^{ABC}).
	\eeax
	Here $E^{A|BC}$ denotes the bipartite measure with respect to bi-partition $A|BC$. 
	In such a case, Eq.~(\ref{power1}) is clear with equality.
	If $z>0$, then there always exists $\gamma>0$ such that 
	\bea \label{f}
	1\leqslant \left(\frac{y}{x}\right)^\gamma+\left( \frac{z}{x}\right)^\gamma
	\eea
	since $\left( \frac{y}{x}\right)^\gamma \rightarrow 1$ 
	and $\left( \frac{z}{x}\right)^\gamma\rightarrow 1$ when $\gamma$ decreases.
	Let $f(\rho^{ABC})$ be the largest value of $\gamma$ for which the inequality~(\ref{f}) is saturated.
	$E$ is continuous leads to $f$ is continuous. 
	Since $f$ is continuous and the set of all pure states in $\mS^{ABC}$ is compact,
	we get
	\be\label{alpha}
	\alpha\equiv\inf\limits_{\rho^{ABC}\in\mS^{ABC}}f(\rho^{ABC})<\infty\;,
	\ee 
	which satisfies Eq.~(\ref{power1}). The proof is completed.
\end{proof}

\begin{figure}
	\hspace{30mm}\includegraphics[width=65mm]{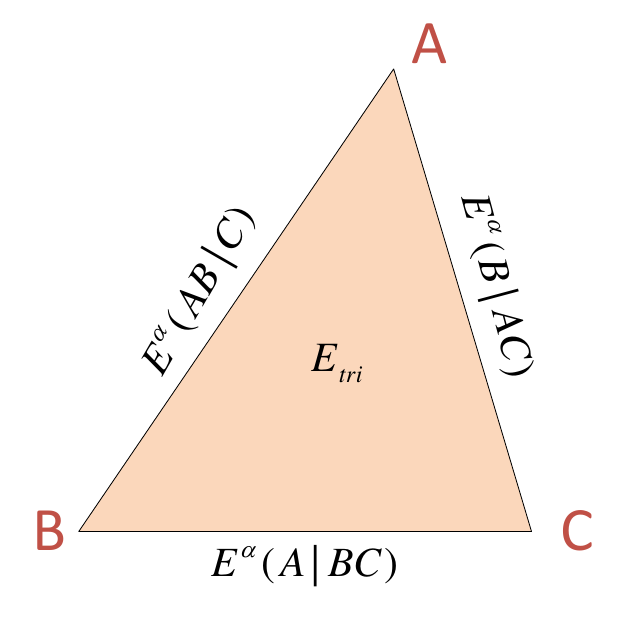}
	%\vspace{2mm}
	\caption{\label{fig1}(color online). The $E$-triangle for a tripartite system. The
		lengths of the three edges are set equal to the $\alpha$-th power of the three
		one-to-other bipartite entanglement measured by $E$. }
\end{figure}

Note that almost all entanglement measures so far are continuous~\cite{GG} and that $E^{\alpha}(|\psi\ra^{A|BC})\leqslant  E^{\alpha}(|\psi\ra^{B|AC})
+ E^{\alpha}(|\psi\ra^{AB|C})$
implies $E^{\gamma}(|\psi\ra^{A|BC})\leqslant  E^{\gamma}(|\psi\ra^{B|AC})
+ E^{\gamma}(|\psi\ra^{AB|C})$
for any $\gamma\in[0,\alpha]$. 
We denote the exponent $\alpha$ associated with $E$ by $\alpha(E)$.
According to Theorem~\ref{plogamy-3} and the arguments in the proof we know that
$E^{\alpha}(|\psi\ra^{A|BC})$, $E^{\alpha}(|\psi\ra^{B|AC})$ and
$E^{\alpha}(|\psi\ra^{AB|C})$ can induce a triangle (the case of $x=y$, $z=0$ is reduced to a line segment which is regarded as a trivial triangle hereafter), which we call it $E$-triangle (See in Fig.~\ref{fig1}).
Going further, we denote the circumradius and the inscribed radius of the $E$-triangle by $R_E$ and $r_{E}$, respectively.
We call the triangle, induced by $E^{\gamma}(|\psi\ra^{A|BC})$, $E^{\gamma}(|\psi\ra^{B|AC})$ and
$E^{\gamma}(|\psi\ra^{AB|C})$ with $0<\gamma<\alpha(E)$, $E$-triangle with exponent $\gamma$.
And we denote the circumradius and the inscribed radius of the $E$-triangle with exponent $\gamma$ by $R_{E(\gamma)}$ and $r_{E(\gamma)}$, respectively.

For any $|\psi\ra\in\mH^{ABC}$,
we write $E^{\alpha}_{A|BC}=x$, $E^{\alpha}_{AB|C}=y$, $E^{\alpha}_{B|AC}=z$ and denote the area of the $E$-triangle of $|\psi\ra$ by $E_{tri}(|\psi\ra)$.
It is worth mentioning that we can not know whether $E_{tri}$
is nonincreasing under LOCC (the area of a triangle could increase even though the edges are decreased~\cite{Cheng}).
We only know $E_{tri}(|\psi\ra)>0$ if and only if $|\psi\ra$ is genuinely entangled. 
We thus turn to consider another quantity 
\bea\label{E-tri}
R_E(|\psi\ra)\cdot r_E(|\psi\ra)=\frac{xyz}{4E_{tri}(|\psi\ra)}\cdot\frac{2E_{tri}(|\psi\ra)}{x+y+z}
=\frac{xyz}{2(x+y+z)},
\eea
whenever $xyz>0$. 
We can verify that $R_E(|\psi\ra)\cdot r_E(|\psi\ra)$ can not increase when
the edges decreasing (in fact, $\frac{\prod_i^mx_i}{\sum_i^mx_i}$ decreases whenever $x_i$s decrease for any $m$, $\sum_i^mx_i>0$).
In addition, in the expression $\frac{xyz}{2(x+y+z)}$, whether $E_{A|BC}$, $E_{AB|C}$, and $E_{B|AC}$ make up a triangle is 
not necessary.
We thus define 
\bea\label{gmem3}
\mF_{123}(\rho)\equiv\left\lbrace \begin{array}{ll}
	\frac{xyz}{x+y+z},&~x+y+z>0,\\
	0,&~x+y+z=0
	\end{array}\right. ~{\rm or}~\mF_{123}(\rho)\equiv{xyz}
\eea
for any $\rho\in\mS^{ABC}$
with $E^{A|BC}(\rho)=x$, $E^{AB|C}(\rho)=y$, $E^{B|AC}(\rho)=z$. 
For any LOCC acting on $\rho$, we assume the output state is $\rho'$. 
Then 
\bea\label{monotonic-3}
\mF_{123}(\rho')\leqslant \mF_{123}(\rho)
\eea
since $E$ is an entanglement measure which implies that
$E^{X|YZ}(\rho')\leqslant E^{X|YX}(\rho)$ for any possible $\{X,Y,Z\}=\{A,B,C\}$.	
It is clear that $\mF_{123}(\rho)=0$ whenever $\rho$ is fully separable (i.e., 3-separable),
and that there exists 2-separable but not fully separable state $\sigma$ such that
$\mF_{123}(\sigma)=0$.
We thus get the following theorem.

\begin{theorem}\label{3-genuine-1}
	Let $E$ be a bipartite entanglement measure.
	Then $\mF_{123}$ is a 3-entanglement measure but not a faithful 3-entanglement measure.
\end{theorem}

\begin{remark}
$\mF_{123}$ is not a genuine entanglement measure in general since there exists
2-separable but not fully separable mixed state $\sigma$ such that
$\mF_{123}(\sigma)>0$ in high-dimension system.	
\end{remark}

It is immediate to realize that 
$\mF_{123}^F$ is not an entanglement measure any more in general
since we can not guarantee that it is nonincreasing under LOCC even for pure state.
When we take
\bea
\mE_{g-123}(\rho)=\delta(xyz)(x+y+z),\\
\mE_{123}(\rho)=x+y+z,
\eea
with 
$x$, $y$, $z$
as above and define
$\mE^F_{g-123}$ and $\mE_{123}^F$ 
as in Eq.~(\ref{eofmin}), it turns out that
$\mE_{g-123}(|\psi\ra)>0$ if and only if
$|\psi\ra$ is genuinely entangled, and $\mE_{123}(\rho)>0$ if and only if
$\rho$ is 3-entangled. Here $\delta(x)=0$ whenever $x=0$, otherwise $\delta(x)=1$.
On the other hand, one can easily check that $\mE_{g-123}$ $\mE^F_{g-123}$, $\mE_{123}$ and $\mE^F_{123}$ are nonincreasing under LOCC and do not increase on average under stochastic LOCC whenever $E$ is an entanglement monotone.
We thus obtain a genuine tripartite entanglement measure.

\begin{theorem}\label{3-genuine-2}
	Let $E$ be a bipartite entanglement measure (resp. monotone).
	Then $\mE^F_{g-123}$ is a genuine tripartite entanglement measure (resp. monotone), $\mE_{g-123}$, $\mE_{123}$
	and $\mE_{123}^F$ are 3-entanglement measures (resp. monotones) but not genuine entanglement measures (resp. monotones).
\end{theorem}
  
\begin{proof}
	The case of $\mE^F_{g-123}$ is obvious according to the argument above.
	For $\mE_{g-123}$, $\mE_{123}$
	and $\mE_{123}^F$, if $\rho$ is fully separable, then $\mE_{123}(\rho)=\mE^F_{123}(\rho)=0$.
	In addition, if $\sigma$ is biseparable but not fully separable, then $\mE_{123}(\sigma)>0$,
	$\mE^F_{123}(\sigma)>0$. 
	When we take
	\beax
	\sigma=p_1|\psi\ra\la\psi|^A\otimes|\psi\ra\la\psi|^{BC}+
	p_2|\phi\ra\la\phi|^{AB}\otimes|\phi\ra\la\phi|^{C}
	+p_3|\xi\ra\la\xi|^B\otimes|\xi\ra\la\xi|^{AC}
	\eeax
	with $|\psi\ra^{BC}$, $|\phi\ra^{AB}$ and $|\xi\ra^{AC}$ being entangled,
	$p_1+p_2+p_3=1$, $p_1p_2p_3>0$,
	then it is possible that $\mE_{g-123}(\sigma)>0$ for some choice of 
	$\{|\psi\ra^A, |\psi\ra^{BC}, |\phi\ra^C, |\phi\ra^{AB}, |\xi\ra^B, |\xi\ra^{AC}\}$.
	The proof is completed.
\end{proof}

We take tangle for example to explain further.
We can easily get 
\bea
\tau(|\psi\ra^{A|BC})\leqslant\tau(|\psi\ra^{AB|C})+\tau(|\psi\ra^{B|AC})
\eea
holds for any $|\psi\ra^{ABC}\in\mH^{ABC}$ 
since~\cite[Theorem 2]{Audenaerta} 
\beax 
1+\tr( \rho^{AB})^2\geqslant\tr( \rho^A)^2+\tr( \rho^B)^2.
\eeax
It is straightforward that the $\tau$-triangle for the GHZ state $\frac{1}{\sqrt{2}}(|000\ra+|111\ra)$ is a regular triangle with side length is 1.
Thus 
\bea
\mF^\tau_{123}(|GHZ\ra)=\frac{1}{3},~{\rm or}~\mF^\tau_{123}(|GHZ\ra)=1,\\
\tau_{g-123}(|GHZ\ra)=\tau_{123}(|GHZ\ra)=3.
\eea
Here, by $\mF^\tau_{123}$ we denote the measure $\mF_{123}$ induce by $\tau$.
For the W state $|W\ra=\frac{1}{\sqrt{3}}(|100\ra+|010\ra+|001\ra)$, the $\tau$-triangle is a regular triangle with side length is $\frac{8}{9}$ and
\bea
\mF^\tau_{123}(|W\ra)=\frac{64}{243},~{\rm or}~\mF^\tau_{123}(|W\ra)=\frac{512}{729},\\
\tau_{g-123}(|W\ra)=\tau_{123}(|W\ra)=\frac{8}{3}.
\eea
Further, since $\tau_{123}(|GHZ\ra)>\tau_{123}(|W\ra)$, and thus $\tau_{123}^F$ is also a ``proper'' genuine tripartite entanglement measure in the sense of Ref.~\cite{Xie2021prl}.

\section{Four-partite entanglement}

In the previous section, we derive $\mF_{123}$, $\mE_{g-123}$ and $\mE_{123}$ from the
bipartite entanglement measure $E$. With this spirit in mind,
moving to four-partite case, 
there are two ways to derive entanglement measures: from bipartite entanglement measure $E$ or
three-partite entanglement measure $E^{(3)}$.

\subsection{From bipartite entanglement measure}

For any given $\rho\in\mS^{ABCD}$ and any bipartite entanglement measure $E$,
we let
\beax
E(\rho^{AB|CD})&=&x_1^{(2)},\\
E(\rho^{A|BCD})&=&x_2^{(2)},\\
E(\rho^{AC|BD})&=&x_3^{(2)},\\
E(\rho^{ABC|D})&=&x_4^{(2)},\\
E(\rho^{AD|BC})&=&x_5^{(2)},\\
E(\rho^{B|ACD})&=&x_6^{(2)},\\
E(\rho^{C|ABD})&=&x_7^{(2)},
\eeax
and define
\bea\label{1234(2)}
\mF_{1234(2)}(\rho)\equiv\left\lbrace \begin{array}{ll}
	\frac{\prod_ix_i^{(2)}}{\sum_ix_i^{(2)}},&~\sum_ix_i^{(2)}>0,\\
	0,&~\sum_ix_i^{(2)}=0
	\end{array}\right. 
	~{\rm or}~
\mF_{1234(2)}(\rho)\equiv\prod_ix_i^{(2)}.
\eea
Then $\mF_{1234(2)}$ is a 4-entanglement measure straightforwardly.
Moreover, 
we let
\bea\label{1234(2)-2}
\mE_{1234(2)}(\rho)\equiv\sum_ix_i^{(2)}
\eea
and 
\bea\label{1234(2)-2}
\mE_{g-1234(2)}(\rho)\equiv\delta\left( \prod_ix_i^{(2)}\right) \sum_ix_i^{(2)}.
\eea
Using the similar argument as that of Theorem~\ref{3-genuine-2}, we can check that $\mE_{1234(2)}$, $\mE_{1234(2)}^F$ and $\mE_{g-1234(2)}$ are 4-entanglement measure
while $\mE_{g-1234(2)}^F$ is a genuine four-partite entanglement measure.
We conclude this argument with the following theorem.

\begin{theorem}
	Let $E$ be a bipartite entanglement measure (resp. monotone).
	Then 
	\begin{itemize}
		\item $\mF_{1234(2)}$, $\mE_{1234(2)}$, $\mE_{1234(2)}^F$ and $\mE_{g-1234(2)}$ are 4-entanglement measures (resp. monotones) but not genuine entanglement measures  (resp. monotones).
		\item $\mE_{g-1234(2)}^F$ is a genuine faithful four-partite entanglement measure (resp. monotone).
	\end{itemize}
\end{theorem}

We illustrate the measures here with $|GHZ_4\ra=\frac{1}{\sqrt{2}}(|0000\ra+|1111\ra)$ and $|W_4\ra=\frac{1}{2}(|1000\ra+|0100\ra+|0010\ra+|0001\ra)$ and the associated bipartite entanglement measure $\tau$.
For $|GHZ_4\ra$, we have $x_i^{(2)}=1$, $i=1$, 2, $\dots$, 7, 
and for $|W_4\ra$, $x_1^{(2)}=x_3^{(2)}=x_5^{(2)}=1$, $x_2^{(2)}=x_4^{(2)}=x_6^{(2)}=x_7^{(2)}=\frac34$.
It turns out that
\bea
\mF_{1234(2)}(|GHZ_4\ra)&=&\frac17~{\rm or}~\mF_{1234(2)}(|GHZ_4\ra)=1,\\
\mE_{1234(2)}(|GHZ_4\ra)&=&\mE_{g-1234(2)}(|GHZ_4\ra)=7,
\eea
and
\bea
\mF_{1234(2)}(|W_4\ra)&=&\frac{27}{512}~{\rm or}~\mF_{1234(2)}(|W_4\ra)=\frac{81}{256},\\
\mE_{1234(2)}(|W_4\ra)&=&\mE_{g-1234(2)}(|W_4\ra)=6.
\eea
Obviously, $|GHZ_4\ra$ is more entangled than $|W_4\ra$ under these measures.

\subsection{From tripartite entanglement}

For any given $\rho\in\mS^{ABCD}$ and any tripartite entanglement measure $E^{(3)}$ (e.g., the unified tripartite entanglement measure as in Ref.~\cite{G2020}, or other tripartite entanglement measure as $\mF_{123}$, $\mE_{123}$, $\mE_{g-123}$, $\mE_{123}^F$, $\mE_{g-123}^F$ discussed in the previous section),
we let
\beax
E^{(3)}(\rho^{A|B|CD})&=&x_1^{(3)},\\
E^{(3)}(\rho^{A|BC|D})&=&x_2^{(3)},\\
E^{(3)}(\rho^{AC|B|D})&=&x_3^{(3)},\\
E^{(3)}(\rho^{AB|C|D})&=&x_4^{(3)},\\
E^{(3)}(\rho^{AD|B|C})&=&x_5^{(3)},\\
E^{(3)}(\rho^{A|BD|C})&=&x_6^{(3)},
\eeax
and define
\bea\label{1234(3)}
\mF_{1234{(3)}}(\rho)\equiv\left\lbrace \begin{array}{ll}\frac{\prod_ix_i^{(3)}}{\sum_ix_i^{(3)}},&~\sum_ix_i^{(3)}>0,\\
	0,&~\sum_ix_i^{(3)}=0,
	\end{array}\right. ~{\rm or}~\mF_{1234{(3)}}(\rho)\equiv\prod_ix_i^{(3)}.
\eea
In addition, let 
\bea \label{3-genuine-vari}
\tilde{E}^{(3)}(\rho^{P|Q|R})=\delta(\cdot|\cdot)E^{(3)}(\rho^{P|Q|R})
\eea
for any three-partition $P|Q|R$ of $ABCD$, $\delta(\cdot|\cdot)=0$ if $\rho^{ABCD}$ is biseparable up to some bi-partition and $\delta(\cdot|\cdot)=1$ otherwise.
Define
\beax
\tilde{E}^{(3)}(\rho^{A|B|CD})&=&\tilde{x}_1^{(3)},\\
\tilde{E}^{(3)}(\rho^{A|BC|D})&=&\tilde{x}_2^{(3)},\\
\tilde{E}^{(3)}(\rho^{AC|B|D})&=&\tilde{x}_3^{(3)},\\
\tilde{E}^{(3)}(\rho^{AB|C|D})&=&\tilde{x}_4^{(3)},\\
\tilde{E}^{(3)}(\rho^{AD|B|C})&=&\tilde{x}_5^{(3)},\\
\tilde{E}^{(3)}(\rho^{A|BD|C})&=&\tilde{x}_6^{(3)},
\eeax
and 
\bea\label{1234(3)'}
\tilde{\mF}_{1234(3)}(\rho)\equiv\left\lbrace 
\begin{array}{ll}\frac{\prod_i\tilde{x}_i^{(3)}}{\sum_i\tilde{x}_i^{(3)}},&~\sum_i\tilde{x}_i^{(3)}>0,\\
	0,&~\sum_i\tilde{x}_i^{(3)}=0,
\end{array}\right. 
{\rm or}~\tilde{\mF}_{1234(3)}(\rho)\equiv{\prod_i\tilde{x}_i^{(3)}}.
\eea
Analogous to $\mE_{g-123}$ and
$\mE_{123}(\rho)$, we define
\bea
\mE_{g-1234(3)}(\rho)=\delta(\cdot|\cdot)\sum_ix_i^{(3)},\\
\mE_{1234(3)}(\rho)=\sum_ix_i^{(3)}.
\eea

It is clear that for any fully separable state $\rho\in\mS^{ABCD}$,
$\mF_{1234(3)}(\rho)=\tilde{\mF}_{1234(3)}(\rho)=\mE_{g-1234(3)}(\rho)=\mE_{1234(3)}(\rho)={\mE}^F_{1234(3)}(\rho)=0$. 
For any pure state $|\psi\ra\in\mH^{ABCD}$,
${\mE}_{g-1234(3)}(|\psi\ra)>0$
if and only if $|\psi\ra$ is genuinely entangled.
In addition, all these quantities do not increase
under any $4$-partite LOCC since $E^{(3)}$ is non-increasing under any 3-partite LOCC.
$\mF_{1234(3)}$, $\tilde{\mF}_{1234(3)}$, $\mE_{1234(3)}$ and $\mE_{1234(3)}^F$ are not GMEMs in general since
these measures are defined
with respect to the fixed partition. For instance,
$\rho^{X|Y|ZW}$s is not separable with respect to partition $X|Y|ZW$ 
can not guarantee $\rho^{ABCD}$ is genuinely entangled.
We can now conclude the following.

\begin{theorem}
	Let $E^{(3)}$ be a tripartite entanglement measure (resp. monotone).
	Then
	\begin{itemize}
		\item $\mF_{1234(3)}$, $\tilde{\mF}_{1234(3)}$,  $\mE_{g-1234(3)}$, $\mE_{1234(3)}$, $\mE_{1234(3)}^F$, and  are 
		4-entanglement measures (resp. monotones).
		\item ${\mE}^F_{g-1234(3)}$ is a faithful genuine four-partite entanglement measure (resp. monotone).
	\end{itemize}
\end{theorem}

Simple calculation reveals that, $\tau^{(3)}(X|Y|ZW)$ are $\frac{3}{2}$ and $\frac{5}{4}$
for $|GHZ_4\ra$ and $|W_4\ra$ respectively for any tripartition of $ABCD$.
It follows that
\beax
\mF^{\tau}_{1234(3)}(|GHZ_4\ra)=\tilde{\mF}{\tau}_{1234(3)}(|GHZ_4\ra)=\frac{81}{64},\\
\mF^{\tau}_{1234(3)}(|W_4\ra)=\tilde{\mF}{\tau}_{1234(3)}(|W_4\ra)=\frac{3125}{1536}
\eeax
or
\beax
\mF^{\tau}_{1234(3)}(|GHZ_4\ra)=\tilde{\mF}{\tau}_{1234(3)}(|GHZ_4\ra)=\frac{27}{8},\\
\mF^{\tau}_{1234(3)}(|W_4\ra)=\tilde{\mF}{\tau}_{1234(3)}(|W_4\ra)=\frac{125}{64},
\eeax
and 
\beax
\mE^{\tau}_{1234(3)}(|GHZ_4\ra)=\tilde{\mF}{\tau}_{1234(3)}(|GHZ_4\ra)=\frac{9}{2},\\
\mE^{\tau}_{1234(3)}(|W_4\ra)=\tilde{\mF}{\tau}_{1234(3)}(|W_4\ra)=\frac{15}{4}.
\eeax
$\frac{27}{8}>\frac{125}{64}$, but $\frac{3125}{1536}>\frac{81}{64}$,
that is, the former measure is not ``proper''.

For any $m$-partite system whenever $m>4$, the scenario above is also feasible.
In fact, for any function $f(x_1,x_2,\cdots, x_n)$ that satisfies
the monotonic condition, i.e.,
\bea
f(x_1',x_2',\cdots, x_n')\leqslant f(x_1,x_2,\cdots, x_n)
\eea
whenever $x_i'\leqslant x_i$, $i=1$, $\dots$, $n$, $n\geqslant m$,
can educe the corresponding entanglement measures.
Moreover, if 
$f$ satisfies 
\bea
f(x_1,x_2,\cdots, x_n)\geqslant\sum_jp_jf(x_1^{(j)},x_2^{(j)},\cdots, x_n^{(j)})
\eea
for any probability distribution $\{p_j\}$ and $\{x_i^{(j)}\}$ such that $x_i\geq \sum_jp_jx_i^{(j)}$,
$f$ can educe an entanglement measure for pure states and then extending to mixed states
via the convex-roof extension.
For the genuine measure, it requires
\bea
f(x_1,x_2,\cdots, x_n)=0
\eea
whenever the state is biseparable.
In general, any $k$-partite entanglement measure $E^{(k)}$ can educe $m$-partite entanglement measures, $2\leqslant k<m$.
All of these multipartite measures can be easily followed analogous to that of four-partite case.

\section{The tetrahedron structure for the four partite state}

For any four-partite state $\rho^{ABCD}\in\mS^{ABCD}$, there are two classes
of quantities: (i) $E^{(3)}(A|B|CD)$, $E^{(3)}(A|BC|D)$, $E^{(3)}(AB|C|D)$,
$E^{(3)}(AC|B|D)$, $E^{(3)}(AD|B|C)$ and $E^{(3)}(A|C|BD)$ derived from any given tripartite 
entanglement measure $E^{(3)}$; (ii) $E(AB|CD)$, $E(AC|BD)$, $E(AD|BC)$ for any given bipartite 
entanglement measure $E$.
In the following, with the same spirit as that of $E$-triangle, we discuss whether these two kinds of quantities can constitute a tetrahedron or
triangle respectively.

\subsection{Tetrahedron based on tripartition}

When it comes to identifying whether six given quantities can constitute a tetrahedron,
we firstly need to check the following triangle relation which can also be regarded as
a polygamy relation of entanglement.

\begin{figure}	
	\hspace{30mm}\includegraphics[width=65mm]{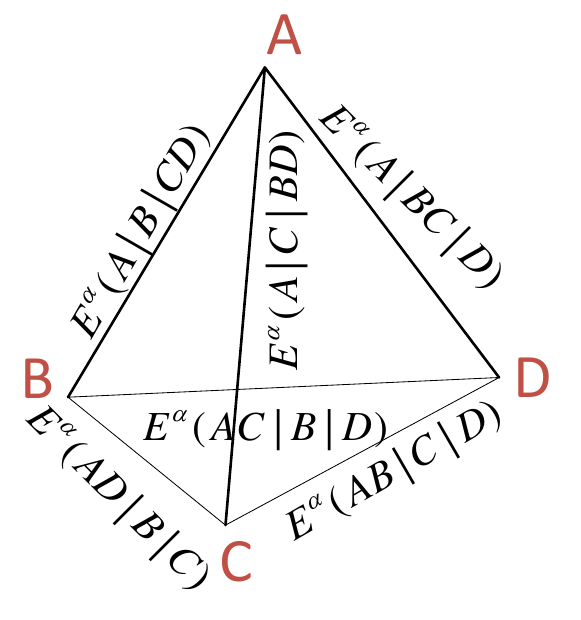}
	\vspace{-4mm}
	\begin{center}
		{Case~(a)}
	\end{center}
	\vspace{-1mm}
	\caption{\label{fig2}(color online). The tetrahedron for a 4-partite system for the case (a). The
		lengths of the 6 edges are set equal to the $\alpha$-th power of $E^{(3)}(|\psi\ra^{X|Y|ZW})$, $\{X,Y,Z,W\}=\{A,B,C,D\}$. }
\end{figure}

\begin{pro}\label{plogamy-4(3.1)}
	Let $E^{(3)}$ be a continuous unified tripartite entanglement measure.
	Then there exists
	$0<\alpha<\infty$ such that	
	\be\label{tetra-condition-1}
	E^{\alpha}(|\psi\ra^{X|Y|ZW})\leqslant  E^{\alpha}(|\psi\ra^{X|YW|Z})
	+ E^{\alpha}(|\psi\ra^{XW|Y|Z})
	\ee
	for all $|\psi\ra^{ABCD}\in\mathcal{H}^{ABCD}$ with fixed $\dim\mH^{ABCD}=d<\infty$, where $\{X|Y|ZW\}$ runs over all possible tripartite partition of $ABCD$, $\{X,Y,Z,W\}=\{A,B,C,D\}$.
	Here we omit the superscript $^{(3)}$ of $E^{(3)}$ for brevity.
\end{pro}

\begin{proof}
	With no loss of generality, we only need to check
	$E^{(3)}(|\psi\ra^{A|B|CD})=E^{(3)}(|\psi\ra^{A|BD|C})$ whenever
	$E^{(3)}(|\psi\ra^{AD|B|C})=0$. The other cases can be followed analogously.
	If $E^{(3)}(|\psi\ra^{AD|B|C})=0$, then 
	$|\psi\ra^{AD|B|C}=|\psi\ra^{AD}|\psi\ra^B|\psi\ra^C$.
	This leads to $E^{(3)}(|\psi\ra^{A|B|CD})=E^{(3)}(|\psi\ra^{A|BD|C})=E^{(2)}(|\psi\ra^{AD})$.
	The proof is completed.	
\end{proof}

\begin{figure}
	%\vspace{3mm}	
	\hspace{30mm}\includegraphics[width=65mm]{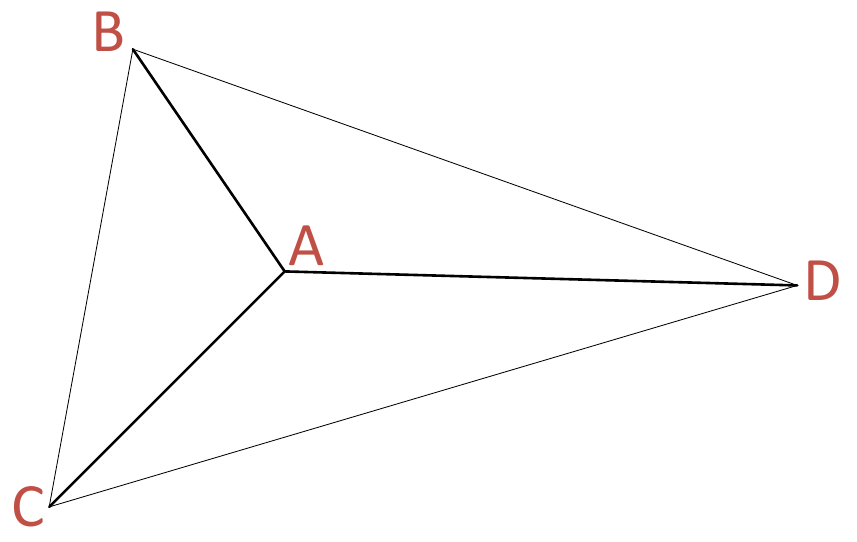}
	\vspace{-2mm}
	\begin{center}
		{Case~(b1)}
	\end{center}
	\vspace{-1mm}
	%\vspace{1mm}
	\caption{\label{fig3}(color online). The triangle structure for a 4-partite system in the case of (b1) (Assume that $S_{BCD}=S_{ABC}+S_{ACD}+S_{ABD}$).}
\end{figure}

\begin{figure}	
	\hspace{30mm}\includegraphics[width=40mm]{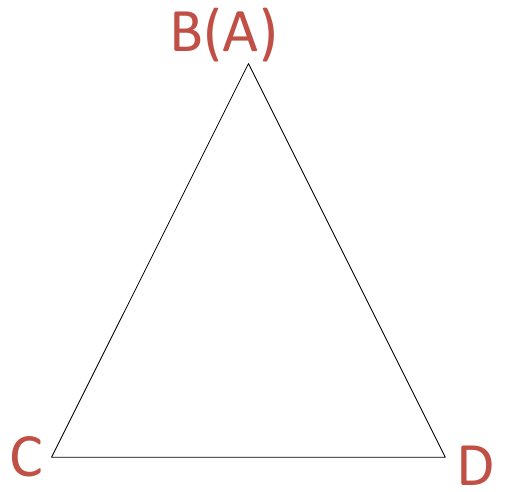}%\vspace{2mm}
	\begin{center}
		{Case~(b2)}
	\end{center}
	\vspace{-1mm}
	%\vspace{1mm}
	\caption{\label{fig4}(color online). The regular triangle structure for a 4-partite system in the case of (b2) (Assume that $E^{(3)}(|\psi\ra^{A|B|CD})=0$ and $E^{(2)}(|\psi\ra^{CD})>0$).}
\end{figure}

\begin{figure}	
	\vspace{3mm}
	\hspace{30mm}\includegraphics[width=65mm]{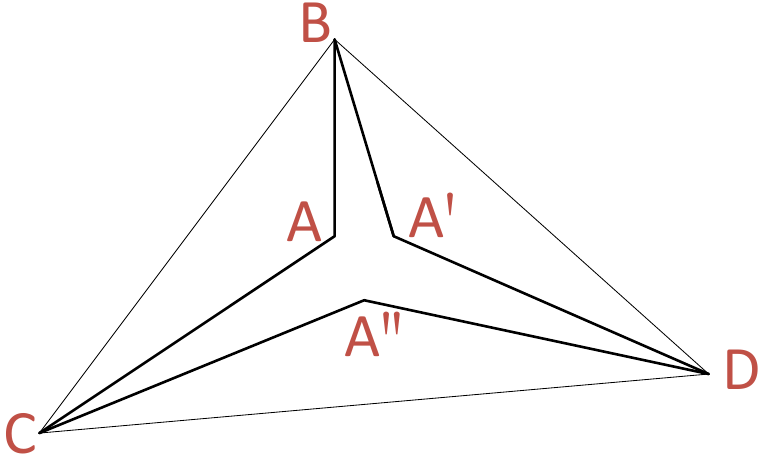}
	\begin{center}
		{Case (c)}
	\end{center}
	\vspace{-1mm}
	\caption{\label{fig5}(color online). The triangle structure for a 4-partite system in the case of (c) (Assume that $E^{(3)}(|\psi\ra^{ABCD})>0$ for any tripartite partition of $ABCD$ and $S_{BCD}>S_{ABC}+S_{ACD}+S_{ABD}$).}
\end{figure}

For any given pure state $|\psi\ra\in\mH^{ABCD}$,
we let $S_{RST}$ denotes the area of the triangle $\bigtriangleup_{RST}$ induced by $E^{\alpha}$ as in Eq.~(\ref{tetra-condition-1}), $R|S|T$ is any three-partition of $ABCD$ (e.g., $AC|B|D$ denoted here by
$R|S|T$ with $R=AC$, $S=B$ and $T=D$ is a three-partition).
According to the proof of Proposition~\ref{plogamy-4(3.1)}, we can conclude that
if $E^{(3)}(|\psi\ra^{X|Y|ZW})=0$ for some $\{X,Y,Z,W\}=\{A,B,C,D\}$,
then $E^{(3)}(|\psi\ra^{R|S|TQ})=E^{(2)}(|\psi\ra^{ZW})$
for all $\{R,S,T,Q\}=\{A,B,C,D\}$ except for 
$X|Y|ZW=R|S|TQ$. That is,
if $E^{(3)}(|\psi\ra^{X|Y|ZW})=0$ for some $\{X,Y,Z,W\}=\{A,B,C,D\}$,
then the three cases above are reduced to
a single regular triangle with edge is $E^{(2)}(|\psi\ra^{ZW})$.
Therefore, there are three different cases:
\begin{itemize}
	\item [(a)] $S_{XYZ}+S_{XYW}+S_{XZW}> S_{YZW}$, for any $\{X,Y,Z,W\}=\{A,B,C,D\}$ (See in Fig.~\ref{fig2}).
	\item [(b)] $S_{XYZ}+S_{XYW}+S_{XZW}= S_{YZW}$ for some $\{X,Y,Z,W\}=\{A,B,C,D\}$ (See in Figs.~\ref{fig3} and ~\ref{fig4}).
	\item [(c)] $S_{XYZ}+S_{XYW}+S_{XZW}< S_{YZW}$ 
	for some $\{X,Y,Z,W\}=\{A,B,C,D\}$ (See in Fig.~\ref{fig5}).
\end{itemize}

That is, 
for the case (a), $S_{RST}>0$ for any three-partition $R|S|T$ of $ABCD$ and $E^{\alpha}(|\psi\ra^{X|Y|ZW})$ can make up a tetrahedron, $\{X,Y,Z,W\}=\{A,B,C,D\}$ (see in (a) of Fig.~\ref{fig2}). We denote by $\mS_p^{ABCD}$
the set of all pure states in $\mS^{ABCD}$, 
and denote the set of all pure states for this case in $\mS_p^{ABCD}$ by $\mS^{ABCD}_{tetra}$. Clearly, $\mS^{ABCD}_{tetra}\subsetneq\mS_p^{ABCD}$.
For the case (b), there are three different subcases:
(b1) $S_{RST}>0$ for any three-partition $R|S|T$ of $ABCD$. 
We denote the set of all states for this subcase in $\mS_p^{ABCD}$ by $\mS^{ABCD}_{\bigtriangleup-1}$.
(b2) $E^{(3)}(|\psi\ra^{X|Y|ZW})=0$ and $E^{(2)}(|\psi\ra^{ZW})>0$ for some $\{X,Y,Z,W\}=\{A,B,C,D\}$.
In such a situation, $E^{(3)}(|\psi\ra^{X|Y|ZW})$s reduces to a regular triangle with edge is $E^{(2)}(|\psi\ra^{ZW})$, $\{X,Y,Z,W\}=\{A,B,C,D\}$ (note that, such a triangle is twice-repeated). We denote the set of all states for this subcase in $\mS^{ABCD}_p$ by $\mS^{ABCD}_{\bigtriangleup-2}$. Evidently,
$\mS^{ABCD}_{\bigtriangleup-2}\subsetneq\mS_p^{ABCD}$.
(b3) $E^{(3)}(|\psi\ra^{X|Y|ZW})=E^{(2)}(|\psi\ra^{ZW})=0$ for some $\{X,Y,Z,W\}=\{A,B,C,D\}$.
In such a situation, $E^{(3)}(|\psi\ra^{X|Y|ZW})$s reduces to a point, i.e., the state is fully separable.
For the case (c), we have $S_{RST}>0$ for any three-partition $R|S|T$ of $ABCD$. We denote the set of all states for this subcase in $\mS_p^{ABCD}$ by $\mS^{ABCD}_{\bigtriangleup-3}$.
We conjecture that, both $\mS^{ABCD}_{\bigtriangleup-1}$ and $\mS^{ABCD}_{\bigtriangleup-3}$
are not empty sets.

As one may expect, the superficial area or the volume of the tetrahedron above really 
represents the entanglement contained in the state. However, whether the superficial area or the volume of the tetrahedron can represent as a measures of
four partite entanglement is unknown since proving or disproving theses quantities are monotonic under LOCC seems a hard work.

\subsection{Triangle based on bipartition}

Observing that, almost all bipartite entanglement measure
are related to a function of the reduced state for pure state~\cite{Vidal2000,GG,G2020}.
For example, $\tau(|\psi\ra^{AB})=2\left( 1-\tr( \rho^A)^2\right) $, $E_f(|\psi\ra^{AB})=S\left( \rho^A\right) $.
These measures are in fact determined by the reduced state for pure states.
We start our with the following proposition, which also reports the polygamy relation 
for the four-partite entanglement. 

\begin{figure}	
	\vspace{1mm}
	\hspace{30mm}\includegraphics[width=53mm]{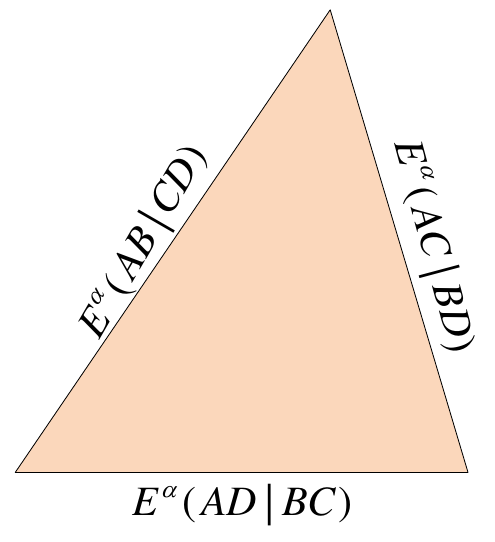}
	\vspace{-1mm}
	\caption{\label{fig6}(color online). The triangle structure for a 4-partite system via bipartite entanglement.}
\end{figure}

\begin{pro}\label{plogamy-4(3.2)}
	Let $E$ be a continuous bipartite entanglement measure that is determined by the eigenvalues of the reduced state.
	Then there exists
	$0<\alpha<\infty$ such that	
	\be\label{tetra-condition-2}
	E^{\alpha}(|\psi\ra^{XY|ZW})\leqslant  E^{\alpha}(|\psi\ra^{XZ|YW})
	+ E^{\alpha}(|\psi\ra^{XW|YZ})
	\ee
	for all $|\psi\ra^{ABCD}\in\mathcal{H}^{ABCD}$ with fixed $\dim\mH^{ABCD}=d<\infty$, where $\{X|Y|ZW\}$ runs over all possible tripartite partition of $ABCD$, $\{X,Y,Z,W\}=\{A,B,C,D\}$.
\end{pro}

\begin{proof}
	With no loss of generality, we only need to verify $E(|\psi\ra^{AB|CD})=0$
	implying $E(|\psi\ra^{AC|BD})=E(|\psi\ra^{AD|BC})$. The other cases are similar.
	If $E(|\psi\ra^{AB|CD})=0$, then $|\psi\ra=|\psi\ra^{AB}|\psi\ra^{CD}$.
	Thus, $E(|\psi\ra^{AC|BD})=h(\rho^{AC})$ and $E(|\psi\ra^{AD|BC})=h(\rho^{AD})$ for some function $h$.
	On the other hand, $\rho^{AC}=\rho^A\otimes\rho^C$, $\rho^{AD}=\rho^A\otimes\rho^D$, and 
	the eigenvalues of $\rho^C$ coincides with that of $\rho^D$ since $|\psi\ra=|\psi\ra^{AB}|\psi\ra^{CD}$.
	It follows that $h(\rho^{AC})=h(\rho^{AD})$ since $h(\rho)$ is only determined by the eigenvalues of $\rho$.
	The proof is completed.
\end{proof}

\begin{figure}	
	\vspace{1mm}
	\hspace{30mm}\includegraphics[width=65mm]{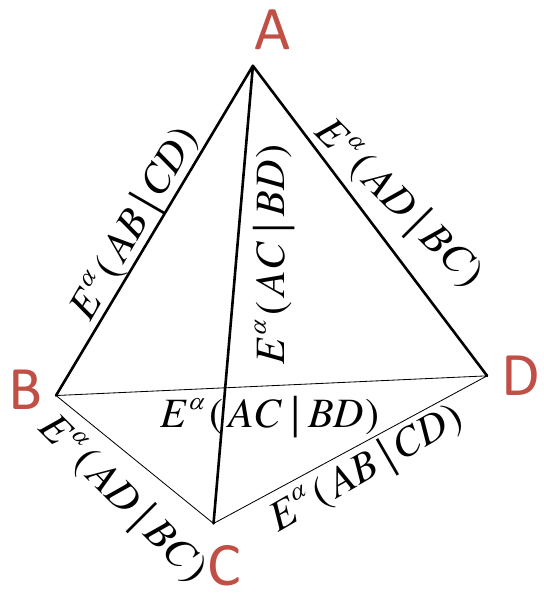}
	\vspace{-1mm}
	\caption{\label{fig7}(color online). The tetrahedron structure for a 4-partite system via bipartite entanglement. The opposite edges are equal.}
\end{figure}

By Proposition~\ref{plogamy-4(3.2)}, any pure state in $\mH^{ABCD}$
can induce a triangle for any bipartite entanglement measure that determined by the reduced state for the pure sates (see in Fig.~\ref{fig6}). In addition,
$E({AB|CD})$, $E({AC|BD})$, $E({AD|BC})$ can be regarded as
6 quantities
$\{E({AB|CD})$, $E({AC|BD})$, $E({AD|BC}), E({CD|AB})$, $E({BD|AC})$, $E({BC|AD})\}$
due to $E(XY|ZW)=E(ZW|XY)$, 
and therefore we can consider whether these 6 quantities can build a tetrahedron.
It depends obviously since in such a case it is equivalent to the fact that whether four same triangles can constitute a tetrahedron (namely, four copies of the same triangle can not necessarily constitute a tetrahedron).
When it does, the tetrahedron is illustrated in Fig.~\ref{fig7}.

Similar to that of $E$-triangle for tripartite case,
whether the area of such a triangle (or, superficial area or volume of tetrahedron if the tetrahedron exists) is a four-partite entanglement measure is unknown.
Going further, it is hard to find a symmetric geometric structure for $m$-partite system whenever $m>4$.
That is, the method in Ref.~\cite{Xie2021prl} is hard to be extended into $m$-partite system whenever $m>3$.

\section{Conclusion and discussion}

In summary, we have established a scenario for establishing genuine entanglement measures and $m$-entanglement measures for tripartite and four-partite quantum system with any
finite dimension of the state space. The method is feasible for $m$-partite system with $m>4$.
The concurrence triangle for three-qubit system
in Ref.~\cite{Xie2021prl} was generalized to any higher dimensional tripartite system and shown to be valid for 
any bipartite entanglement measure.
Although the area of the triangle and the superficial area of the tetrahedron can increase 
whenever the lengths of the edges decrease, whether such area is nonincreasing under LOCC is unknown.
If it is nonincreasing under LOCC, then the area is a well-defined GMEM.
In addition, Eqs.~(\ref{power1}), (\ref{tetra-condition-1}), and (\ref{tetra-condition-2}) indeed report
the polygamy relation of the bipartite entanglement in the multipartite state.
This relation combined with the monogamy relation reveals the distribution
of multipartite entanglement more efficiently.

%-------------------------------------------------------------%

\ack{
This work is supported by the National Natural Science Foundation of
China under Grants No.~11971277 and No.~11901421, the Scientific Innovation Foundation of the Higher 
Education Institutions of Shanxi Province under Grants No.~2019KJ034 and No.~2019L0742,
and the Science Technology Plan Project of Datong City under 
Grant No.~2020155. 
}	
%\section*{Acknowledgements}

%\nocite{*}
%\bibliographystyle{apsrev4-1}
%\bibliography{gy0705}% Produces the bibliography via BibTeX.

%\if false

%\nocite{*}
%\bibliographystyle{apsrev4-1}
%\bibliography{gy0705}

\end{document}